\documentclass[conference]{IEEEtran}
\IEEEoverridecommandlockouts
\usepackage{cite}
\usepackage{amsmath,amssymb,amsfonts}
\usepackage{algorithmic}
\usepackage{graphicx}
\usepackage{textcomp}
\usepackage{xcolor}
\def\BibTeX{{\rm B\kern-.05em{\sc i\kern-.025em b}\kern-.08em
    T\kern-.1667em\lower.7ex\hbox{E}\kern-.125emX}}

\usepackage{amsmath, amssymb, amsthm, amsfonts,cite,alltt,clrscode}

{\catcode`\^^M=\active \gdef\tablines{\catcode`\^^M=\active \def^^M{\\}}}
\newenvironment{pcode}{\tablines \topsep=0pt \partopsep=0pt \tabbing
   MM\=MM\=MM\=MM\=MM\=MM\=MM\=MM\=MM\=MM\=MM\=MM\=MM\=\kill}
  {\endtabbing\vspace*{-0\baselineskip}}
{\makeatletter \gdef\lasttab{\ifnum \@curtab<\@hightab \>\lasttab\fi}}

\begin{document}

\title{An Efficient Reversible Algorithm \\ for Linear Regression
\thanks{Supported by NSERC.}
}

\author{
%   Erik D. Demaine%
%     \thanks{Computer Science and Artificial Intelligence Laboratory,
%       Massachusetts Institute of Technology, 32 Vassar St.,
%       Cambridge, MA 02139, USA,
%       \protect\url{{edemaine,virgi,nwein}@mit.edu}}
\IEEEauthorblockN{ Erik D. Demaine}
\IEEEauthorblockA{\textit{Computer Science and  } \\
\textit{Artificial Intelligence Laboratory, } \\
\textit{Massachusetts Institute of Technology, }\\
Cambridge, MA, USA \\
\protect{edemaine@mit.edu}}
\and
\IEEEauthorblockN{ Jayson Lynch}
\IEEEauthorblockA{\textit{Cheriton School of Computer Science, } \\
\textit{University of Waterloo, }\\
Waterloo, ON, Canada \\
\protect{jayson.lynch@uwaterloo.ca}}
\and
\IEEEauthorblockN{Jiaying Sun}
\IEEEauthorblockA{\textit{Miss Porter's School} \\
Farmington, Connecticut, USA}
}

\maketitle

\begin{abstract}
This paper presents an efficient reversible algorithm for linear regression, both with and without ridge regression. Our reversible algorithm matches the asymptotic time and space complexity of standard irreversible algorithms for this problem. Needed for this result is the expansion of the analysis of efficient reversible matrix multiplication to rectangular matrices and matrix inversion.
\end{abstract}


\section{Introduction}

Machine learning has become a major tool across many domains and part of the increasing success and capabilities can be attributed to more data and computing power. The energy demands of computing have been exponentially increasing despite exponential increases in energy efficiency \cite{IEA, koomey2010implications}. Reversible computing is a technology that has the potential to drastically reduce the energy consumption of computers and is a necessary technology for continued efficiency gains in the long term\cite{frank2005introduction}. However many theoretical questions remain open about reversible computing and many engineering challenges stand in the way of taking advantage of these potential gains. By studying reversible algorithms for machine learning problems we gain a better understanding of theoretical aspects of logical reversibility which we hope will aid future work leading to greener computing. 

%\paragraph*{Results. }
In this paper we give asymptotically efficient algorithms for solving systems of linear equations and for computing linear regressions. Section~\ref{sec:Rectangular Matrix Multiplication} extends the analysis of reversible matrix multiplication to apply to rectangular matrix multiplication. Section~\ref{sec:Matrix Inversion} gives an efficient reversible algorithm for matrix inversion. Section~\ref{sec:Ordinary Least Squares} uses the prior results to give an efficient reversible algorithm for ordinary least squares regression and Section~\ref{sec:Regularizing linear regression} does the same for ridge regression.
\subsection{Related Works}
Although there is a limited number of previous studies on reversible computing for machine learning, there is work on reversible algorithms. We now describe some of the fascinating work on reversible algorithms more broadly. 

\paragraph*{Reversible Neural Networks} Reversible or invertible machine learning algorithms have also been studied without the context of reversible computing. For example algorithms for invertible neural nets have been developed to try to save on the memory requirements for training various types of neural nets\cite{al2004reversible,gomez2017reversible, maclaurin2015gradient,chang2018reversible,mackay2018reversible}. These have gone on to find other uses including the development of involutive neural nets and other architectures which enforce certain group symmetries\cite{spanbauer2020neural}. Although the level of reversibility in these schemes does not suffice for the needs of reversible computing, they certainly make things much easier by ensuring large logical portions of the algorithm are bijective. Hopefully techniques in reversible algorithm design can inspire further techniques.

\paragraph*{Universal Transforms} 
The first universal transforms for reversible computing were independently discovered by Lecerf and Bennett \cite{lecerf63}\cite{bennett73} who make functions bijective by storing inputs at every computation step giving a time complexity $T'(n) = O(T(n))$ and a space complexity $S'(n) = O(S(n)T(n)).$ In 1979, Bennett\cite{bennett73} developed a recursive recomputing, which involves recursively computing to the midpoint of the computation, storing it, and uncomputing to the last checkpoint. The time complexity for this would be $T'(n) = O(T(n)lg(T(n)))$ and space complexity as $S'(n) = O(S(n)lg(T(n))).$ Lange, Mckenzie, Tapp in 2000 \cite{Lange2000ReversibleSE} gave an algorithm which walked the entire computation tree and is reminiscence of Savitch’s Algorithm giving a time complexity of $T'(n) = O(2^{T(n)})$ and a space complexity of $S'(n) = O(S(n)).$ These techniques were combined by Williams in 2000 and Buhrman, Li, Tromp, and Vitanyi \cite{Buhrman97kolmogorovrandom} in 2001 to give a trade-off relationship between time and space. By embedding configuration space enumeration at the bottom of a Bennett recursion they give a family of algorithms for any parameter $k$ in which the time complexity would be $T'(n) = O(S(n)k(T(n)/2^k))$ and time complexity would be $S'(n) = O(kS(n)).$ Xu in 2015 \cite{DBLP:journals/corr/Xu15e} investigated computing with 'dirty ancilla bits' and determined that space can be temporarily reused without knowing its prior state giving a time complexity of $T'(n) = O(2^{T(n)})$ and a space complexity of $S'(n) = S(n)+O(1)$. The study of efficient reversible algorithms for specific problems (as opposed to these universal transforms) has been a more recent area of study and some of this work is described in the following paragraphs.

\paragraph*{Energy Complexity of Algorithms}
Demaine et al.~\cite{demaine2016energy} studied energy variations of computations on circuit RAM, word RAM, and trans-dichotomous RAM. Their paper is the first that performs a systematic study of time/space/energy complexity of reversible computing. In contrast to most previous investigations, they allow algorithms to destroy bits. They developed a high-level pseudocode for fundamental models, such as control logic, memory allocation, garbage collection, and logging and unrolling, and demonstrated that these programming structures may be implemented with zero energy overhead in time and space. Li et al.~\cite{Li1996ReversibilityAA} also investigated time, space, and erasure complexity of computation from an information theoretic viewpoint. They proved in their paper that Bennett's pebbling strategy uses the least space complexity for the greatest efficiency in the pebbling model. Comparing a restatement in terms of $E^{t}(\cdot)$ as $$E^{t'}(x,\in)\le{C^t(x)}+2C^t(t|{x})+4\log C^t(t|{x})$$ with $t'(|x|) = O(2^{|x|}t(|x|)$ to the righthand inequality of equation $E^{t}(x,\in)\geq C^t(x) \geq \frac{1}{2} E^{t'}(x,\in)$, they have improved the upper bound on erasure cost. They also established a trade-off hierarchy of time versus irreversibility for exponential time computations: 
\begin{quote}
    For every large enough $n$ there is a string $x$ of length $n$ and a sequence of $m = \frac{1}{2}\sqrt{n}$ (exponential) time functions $t_{1}(n)<t_{2}(n)<\cdots<t_{m}(n),$ such that 
    $$
    E^{t_1}(x,\in)>E^{t_2}(x,\in)>\cdots>E^{t_m}(x,\in).
    $$
\end{quote}

\paragraph*{Comparison Sorts}
Axelsen et al.~\cite{ComparisonSorts} proposed the following criteria for determining optimality of reversible simulation: A reversible simulation is faithful if it incurs no asymptotic time overheard and bounds the space overhead (the garbage) by some function $g(n),$ and they also showed how to use programming approaches to create reversible simulations of many wells that are both accurate and hygienic (minimizing garbage bits) of several comparison sorting algorithms, including insertion sort and quicksort. They also included prior works on reversible sorting: In $MOQA$ quicksort, there appears a rank (a number in the range $[0, n! - 1]$ indicating the permutation) for demonstrating injectivity. Yokoyama et al.\ used rankings to generate a reversible insertion and merge sorts \cite{Yokoyama}. Lutz et al.\ used a permutation approach to attempt for a reversible bubble sort \cite{Lutz86}. Sorting was also considered by Frank \cite{Frank99} and Demaine et al.~\cite{demaine2016energy}.

\paragraph*{Shortest-Path Algorithms}
Guo et al.~\cite{ShortestPathsProblems} presented several reversible algorithms along with code for shortest paths problems. These include reversible Bellman-Ford algorithm, reversible Dijkstra's algorithm, and reversible Floyd-Marshall algorithm. Most of the reversible programs they have examined have asymptotically minimal additional space, and therefore the resulting time complexity is the same as their irreversible counterparts. 

Frank’s thesis \cite{Frank99} examined sorting algorithms, which he discovered that any standard sorting algorithms have $\Theta(n \log n)$-time comparison, and are easy to turn into good reversible algorithms by saving away bits giving the result of each comparison. Moreover, he also investigated the Floyd-Warshall algorithm which takes $\Theta(n^3)$ time, $\Theta(n^2)$ space. He also examines a reversible iterated matrix multiplication algorithm giving reversible all-pairs shortest path in $\Theta(n^3 \log n)$ time and $\Theta(n^2 \log n)$ space. 

Demaine et al.~\cite{demaine2016energy} also examined Bellman-Ford and Floyd-Warshall algorithms which they found that Reversible Bellman-Ford algorithm runs in $\Theta(VE)$ time, $\Theta(VE)$ space, and $0$ energy and the Reversible Floyd-Warshall algorithm runs in $\Theta(VE)$ time, $\Theta(VE)$ space, and $0$ energy. 

Implementations of many graph algorithms in Janus were provided in \cite{sarah2015reversible}.

\section{Reversible Matrix Algorithms}
\label{Sec:Reversible Matrix Algorithms}

In this section we extend Frank's reversible matrix multiplication algorithm to rectangular matrices (Section~\ref{sec:Rectangular Matrix Multiplication}) and we give an efficient reversible algorithm for computing the inverse of a matrix and for matrix inversion (Section~\ref{sec:Matrix Inversion}). Matrix multiplication, transpose, and inverse will be the primary components needed to perform linear regression.

\subsection{Rectangular Matrix Multiplication}
\label{sec:Rectangular Matrix Multiplication}

Efficient reversible algorithms for square matrix multiplication are discussed in \cite{Frank99} and \cite{demaine2016energy}. The same tricks suffice for making an efficient reversible algorithm for rectangular matrix multiplication.

Given an $m\times{n}$ and a $p\times{n}$ matrix we wish to compute the resulting $p\times{m}$ matrix in time:

$$T(MM(n,m,p)) = O(nmp)$$

and space:
	
$$S(MM(nmp)) = O(mn+pn+pm)$$ 

The standard algorithm runs through all entries of the output matrix and calculates a dot product between the associated vectors. If we use either of the standard universal transforms for reversible algorithms we will either end up with $O(nmp)$ space or a logarithmic overhead in space and time. However, we can treat the innermost dotproduct as a reversible subroutine to save space.

\begin{pcode}
$\proc{RegularMatrixMultiplication}()$: \+
\id{for} $i=1$ \id{to} $m$:    \+  
    \id{for} $j=1$ \id{to} $p$:    \+  
        \id{for} $k=1$ \id{to} $n$:    \+  
			$C_{i,j} \mathrel{{+}{=}} A_{i,k} * {B_{k,j}}$ \
\end{pcode}

\vspace{3mm}

Given a reversible algorithm which calculates some function $f$, we can calculate the output of the function, copy it into another location in memory, then uncompute the function. To remove the output of the reversible subroutine we will need to recalculate the output. Thus at a cost of doubling the time needed we can reduce the storage needed for intermediary calculation to the size of their inputs and outputs. This technique has been observed in some fashion at least since \cite{space} and has been used whether explicitly or implicitly in most efficient reversible algorithms.

%If we want to do reversible computing for matrix multiplication, we’d need to add a temporary value where we can store the intermediate values $A_{i,n}*B_{n,j}$ into the temporary variable temp\_mult. The forward process is just adding a temporary value into the process, which would be different from the regular computation. 
We give pseudocode below which considers the innermost for-loop a reversible subroutine. The individual multiplications of the dot product, $A_{i,n}*B_{n,j}$, are stored into $temp\_mult$ and then added into $C_{i,j}$. After each multiplication we can recompute the multiplication freeing the temporary stored value. For the backwards computation, the dot product is recalculated and stored into $temp\_mult,$ where $C_{i,j}$ is zeroed by subtracting off each element of the dot product. This whole process takes up less space as $O(pm)$ because now we don’t have to waste extra space to store the intermediate values. However, although the multiplication is done four times we maintain the same asymptotic running time of $O(nmp)$.% because it is a constant factor and asymptotic, and for linear computations, the time complexity is not affected even if the computation process is done for more times. 

\begin{pcode}
$\proc{ReversibleMatrixMultiply}()$: \+
Forwards: 
\id{for} $i=1$ \id{to} $m$:    \+  
    \id{for} $j=1$ \id{to} $p$:    \+  
        \id{for} $k=1$ \id{to} $n$:    \+  
			$(temp\_mult, A_{i,k}, B_{k,j}) =$ \\ $(A_{i,k}*B_{k,j}, A_{i,k}, B_{k,j})$ \# forwards
			$C_{i,j} \mathrel{{+}{=}} temp\_mult$
			$temp\_mult \mathrel{{-}{=}} A_{i,k}*B_{k,j}$ \# backwards \-\-\-
			
Backwards:
\id{for} $i=1$ \id{to} $m$:    \+  
    \id{for} $j=1$ \id{to} $p$:    \+  
        \id{for} $k=1$ \id{to} $n$:    \+  
			$(temp\_mult, A_{i,k}, B_{k,j}) =$ \+
    			$(A_{i,k}*B_{k,j}, A_{i,k}, B_{k,j})$
    			\# recalculate $A_{i,k}*B_{k,j}$ \-
			$C_{i,j} \mathrel{{+}{=}} temp\_mult$
			$temp\_mult \mathrel{{-}{=}} A_{i,k}*B_{k,j}$ \# reverse
\end{pcode}

\subsection{Gaussian Elimination and Matrix Inversion}
\label{sec:Matrix Inversion}

In this section we give a reversible version of Gaussian Elimination for computing the inverse of a matrix. One might hope the same technique for matrix multiplication will simply work with a standard matrix inversion algorithm. However, that is not the case and we will briefly examine the issue with Gaussian Elimination before giving a modification to that algorithm which will admit an efficient reversible subroutine and adapt into an efficient reversible algorithm. Pseudocode for the modified Gaussian Elimination is given below.

Gaussian elimination uses row operations to transform a matrix into the Identity. If these same operations are applied to an identity matrix, a vector of unknowns, or other entities this method can be used to solve systems of linear equations or calculate matrix inverses or LU decompositions\cite{clrs}. A basic version of the algorithm starts by looping over all rows in the matrix, normalizing  the first non-zero entry, and then subtracting multiples of that row from all the other rows to zero out all but one entry in a column of the matrix. If we look at the innermost loop, subtracting a multiple of one row from another, we have a sub-routine which is called $O(n^2)$ times and writes over an entire row in the matrix. Since its output space is $O(n)$ we would still need to store a total of $O(n^3)$ space to reverse all of these operations. If we move outward and consider each row zeroing a column of the matrix, we have $O(n)$ calls to a subroutine that writes over the entire matrix. Thus the output space is $O(n^2)$ for each call of the subroutine and we again do not save any space. For matrix multiplication we were able to save because the output space of the dot product was a single number, much smaller than the total running time of the subroutine.

\begin{pcode}
$\proc{Modified Gaussian Elimination}(A)$ \+
    \# Identity matrix which will be modified into the inverse
    Inverse\_$A$ = Identity(size($A$))
    \# Loop over all rows
    \id{for} $i=1$ \id{to} $n$:    \+
        \# Loop over all prior rows
        \id{for} $j=1$ \id{to} $i$:    \+
            \# Calculate multiplier needed to zero out entry $j$
            row\_multiplier $= A_{i,j} / A_{j,i}$
            \id{for} $k=1$ \id{to} $n$:    \+
                \# Write over entries \textbf{\boldmath in row $i$}
                $A_{i,k} \mathrel{{-}{=}} A_{j,k}$ $*$ row\_multiplier
                Inverse\_$A_{i,k} \mathrel{{-}{=}} A_{j,k}$ $*$ row\_multiplier \- \-
        \# normalize based on entry $i$
        reducing\_multiplier $= A_{i,i}$
            \id{for} $k=i$ \id{to} $n$:    \+
                $A_{i,k} = A_{i,k}$ $/$ reducing\_multiplier
                Inverse\_$A_{i,k} = A_{i,k}$ $/$ reducing\_multiplier \-
                
    \# Loop backwards over all rows
    \id{for} $i=n$ \id{to} $1$:    \+
    \# Loop over all prior rows
        \id{for} $j=i+1$ \id{to} $n$:    \+
            \# Calculate multiplier needed to zero out entry $j$
            row\_multiplier $= A_{i,j}$
            \id{for} $k=1$ \id{to} $n$:    \+
                \# Write over entries \textbf{\boldmath in row $i$}
                $A_{i,k} \mathrel{{-}{=}} A_{j,k}$ $*$ row\_multiplier
                Inverse\_$A_{i,k} \mathrel{{-}{=}} A_{j,k}$ $*$ row\_multiplier \- \- \-
    \id{return} Inverse\_$A$

\end{pcode}

Now we describe the key idea in the modified version of the algorithm. Again just considering the first part of the algorithm, converting the matrix to row echelon form. To solve this problem we will reorder the algorithm so that in each loop we have a single row calculate all the updates it needs, thus making changes to only one row on each iteration. In this case we have a subroutine that is still called $O(n)$ times and performs $O(n^2)$ operations, but has an output of size $O(n)$ allowing us to save garbage bits by recalculating. One can think of this as each row pulling information from $O(n)$ other rows to calculate the effect from every other row of the matrix on itself, as opposed to the standard algorithm where each row pushes the information of its impact to every other row.

The second half of the algorithm again goes over all the rows using the previous reduced rows to cancel out remaining entries in the row. Again we make all updates to the same row in the same loop so we have a reversible subroutine with output size $O(n)$ and time $O(n^2)$ which is called $O(n)$ times. After this transformation has finished the program returns the new inverse matrix.

We have not addressed pivoting or other techniques to improve numerical stability or efficiency. Common techniques such as swapping rows based on zero valued entries or the size of the leading entries now cannot occur because partially reduced rows cannot be compared since each row goes form being unchanged to fully reduced before any other rows are altered. Addressing these techniques as well as others used to make matrix inversion more efficient and robust are important challenges moving forward.

\section{Linear Regression}
\label{sec:Linear Regression}
%Regression [] needs a loss function to evaluate the accuracy of our hypothesis by comparing the target value to the predicted value. Square error$(SE)$ allows us to estimate how much the loss is: 
%$$Loss(guess, actual) = (guess - actual) ^2 $$ 

Linear regression is a predictive model with a linear hypothesis class.\footnote{We follow the notation used in \cite{6036}.}
$$h(x; \theta, \theta_0) = \theta^Tx+\theta_0, $$

Typically one minimizes the squared error, so least squares linear regression wants to minimize the following loss function with respect to $\theta$.
$$J(\theta, \theta_0) = \frac{1}{n}\sum^n_{i=1}\left(\theta^Tx^{(i)}+\theta_0-y^{(i)}\right)^2$$
 where $y(i)$ is the actual value and $\theta^Tx^{(i)} + \theta_0$ is the predicted value. This can be solved analytically, and we give an efficient reversible algorithm in Section~\ref{sec:Ordinary Least Squares}. In Section~\ref{sec:Regularizing linear regression} we give an efficient reversible algorithm for ridge regression, a common form of regularized linear regression.

\subsection{Ordinary Least Squares}
\label{sec:Ordinary Least Squares}

Ordinary least squares, or linear least squares, estimates the parameters in a regression model by minimizing the sum of the squared losses. We can achieve this by finding a closed-form formula and we can first set the derivative of $J$ to $0$, and solve for $\theta$. When the derivative of $J$ equals to $0$, $J$ can be either the minimum or maximum, so we have to make sure the point we are calculating is not the maximum or an inflection point. 

Training data can be expressed in the form of matrices. Each column of $X$ is a measure and each $y$ is a predicted data, where $d$ is the dimension of the matrices and $n$ is the number of $x$ columns. 

$$X = \begin{bmatrix}
x_1^{(1)} & \dots & x_1^{(n)}\\
\vdots& \ddots & \vdots \\
x_d^{(1)} &\dots& x_d^{(n)}
\end{bmatrix};
\hspace{2mm} Y = \begin{bmatrix}
y^{(1)} &\dots& y^{(n)}
\end{bmatrix}. $$

We can also transpose the matrices where columns become rows, rows become columns, and the original $y$ horizontal vector now becomes a vertical one. 

$$W = X^T = \begin{bmatrix}
x_1^{(1)} &\dots & x_d^{(1)}\\
\vdots& \ddots & \vdots \\
x_1^{(n)} & \dots & x_d^{(n)} \\
\end{bmatrix};
\hspace{2mm} 
T = Y^T = \begin{bmatrix}
y^{(1)} \\
\vdots\\
y^{(n)}
\end{bmatrix}. $$

Now, we can get

\begin{align*}
(\theta) &= \frac{1}{n}\underbrace{(W\theta-T)^T}_{1\times{n}}\underbrace{(W\theta-T)}_{n\times{1}} \\ 
&=  \frac{1}{n}\sum^n_{i=1}\left(\left(\sum^d_{j=1}W_{ij}\theta_j\right)-T_i\right)^2.
\end{align*}

After setting to $0$ and solving, we get 

$$\theta = \underbrace{(W^TW)^{-1}}_{d\times{d}} \underbrace{W^T}_{d\times{n}}\underbrace{T}_{n\times1}$$
where $\theta$ is the model parameter. 
At the end, we have $\theta$ equal to a specific number because the dimensions can multiply out since the resulting dimension of $(d\times{d})(d\times{n})(n\times{1})$ is $1$. The process should approximately take $O(dn+n^2)$ as time complexity. 
Note, it is possible for $W^TW$ to be singular and this should be addressed if it is relevant to the application area.

Given ordinary least squares regression with $n$ data points of dimension $d,$ $OLS(n,d),$ we will analyse its time and space complexity. In the algorithm we perform one matrix inversion of a $(d\times{d})$ matrix and three matrices multiplies of sizes: $(d\times{n})$ times $(n\times{d})$, $(d\times{d})$ times $(d\times{n})$, $(d\times{d})$ times $(n\times 1)$. This gives a total running time of 
$$T(OLS(n,d)) = O(dn^2+d^3+d^2n).$$ 

If we take the common assumption that our number of data points $n$ is larger than the dimension of the data $d$ then this simplifies to
$$T(OLS(n,d)) = O(dn^2).$$

Similarly we can compute the space complexity given multiple calls to rectangular matrix multiply resulting in
$$ S(OLS(n,d)) = O(d^2+dn) = O(dn).$$

\paragraph*{Reversible Analysis} As we just saw, linear least squares can be expressed as three matrix multiplications and a matrix inverse. Given our efficient reversible rectangular matrix multiplication and matrix inversion from the previous section, we can use that as a reversible subroutine to obtain a reversible ordinary least squares algorithm with the same asymptotic running time and space as the irreversible version. %Given an efficient reversible algorithm one can run that algorithm, copy the output, undo the algorithm recovering the extra space needed and thus obtain the result at a cost of twice the time complexity and an additive factor of space equal to the size of the output. 

If we had used one of the naive transforms, such as Bennett or Lecerf, then the reversible algorithm would have required either $\Theta(d^2n)$ space or suffered a logarithmic overhead in time. Thus, we can compute the linear regression that minimizes the losses.

\subsection{Regularizing linear regression}
\label{sec:Regularizing linear regression}

Regularization is a technique used to reduce errors by fitting the function appropriately and prevent overfitting. It is also possible for $(W^{T}W)$ to be not be invertible. Both of these issues can be addressed with ridge regression which adds an $L_2$-norm regularizer, $\|\theta\|^2$, to the ordinary least squares with a tuning parameter $\lambda$.% which controls the coefficients. As $\lambda$ gets larger even to $\infty$, the coefficients would shrink to zero, where $\|\theta\|^2$ is the squared distance from the predicted value to the actual value. This regularization technique prevents overfitting because it discourages learning a relatively complex model. As for now, the ridge regression objective function equals: 
This gives the following objective function:
$$J_{ridge} = (\theta, \theta_0) = \frac{1}{n}\sum^n_{i=1}(\theta^Tx^{(i)} = \theta_0 - y^{(i)})^2 + \lambda\|\theta\|^2.$$

%This expression of $\theta_0$ minimizes $J_{ridge}$. After composing the same process we have done for $OLS$, setting $J_{ridge}$ to zero and calculating $\theta_{ridge}$, we get
This minimization also has an analytic solution given by
$$\theta_{ridge} = (W^T W + n\lambda{I})^{-1}W^T T,$$
% which is guaranteed to be invertible when the tuning parameter $\lambda$ is greater than $0$. 

This differs from the prior example by scalar multiplications and a matrix addition which do not change the time or space complexity of the reversible or irreversible algorithms.

\section{Conclusion and Open Problems}
\label{sec:Conclusion}
\label{sec:Open Problems}

There are significant engineering challenges that must be overcome to unlock the potential of general purpose reversible computing. These exist at every layer, from physics and gate design through theoretical algorithms. This paper gives asymptotically efficient reversible algorithms for several fundamental linear algebra and machine learning problems taking away another barrier to our goal. In addition, the study of efficient algorithms may help guide the development of special purpose hardware and heterogeneous architectures which use principals from reversible computing to save energy.

Despite significant progress in the past decade, the field of efficient reversible algorithms is still new and many fundamental algorithms have not been studied in this context. This includes essentially the entire field of machine learning. Clustering algorithms, nearest neighbor algorithms, logistic regression, and deep neural nets all seem like compelling cases. Going further one can make more demanding requests of the efficiency of reversible algorithms. For example, our algorithms are not hygienic as they produce more garbage bits than strictly necessary\cite{ComparisonSorts}.

Even within the context of linear regression and linear algebra theoretical questions still abound. In this paper we chose simple, textbook algorithms to start building up more techniques for designing efficient reversible algorithms. One could consider fast matrix multiplication algorithms, numerically stable matrix inversion, cache-oblivious algorithms or the many other techniques used to improve performance both in theory and in practice. Cache-oblivious matrix multiplication\cite{frigo1999cache} and Strassen's algorithm\cite{strassen1969gaussian} do not have a linear size inner loop to use as a reversible subroutine and thus the technique here does not immediately adapt.

Our work here is also purely theoretical. It would be good to see more implementations of reversible algorithms in reversible programming languages like Janus. Further work simulating the resource use of these algorithms or estimating their efficiency on plausible reversible hardware would be useful for understanding the trade-offs of potential reversible computers.

% \section*{Acknowledgment}
% Special thanks to Erik Demaine who suggested the main idea for adapting the Gaussian Elimination algorithm.

%\section*{References}
\bibliographystyle{IEEEtran}
\bibliography{bibliography}

% \section{Appendix}

% def new_invert(A):
%   inverse_A = np.identity(len(A))
%   # iterate through all rows updating them
%   for r in range(len(A)):
%     #search through prior rows and update
%     for k in range(r):
%       # print("k = " + str(k)+ " r = " + str(r))
%       # print("A[r][k] = " + str(A[r][k])+ " A[k][k] = " + str(A[k][k]))
%       row_multiplier = copy.deepcopy(A[r][k]/A[k][k])
%       # print("row_multiplier = " + str(row_multiplier))
%       #update each column entry
%       for c in range(len(A[0])):
%         # print("row_multiplier = " + str(row_multiplier))
%         # print("c = "+ str(c))
%         A[r][c] -= A[k][c]*row_multiplier
%         inverse_A[r][c] -= inverse_A[k][c]*row_multiplier
%     #reduce coefficent to 1
%     row_reducer = copy.deepcopy(A[r][r])
%     for c in range(len(A[0])):
%       A[r][c] = A[r][c]/row_reducer
%       inverse_A[r][c] = inverse_A[r][c]/row_reducer
%     # print("changes")
%     # print(A)
%     # print(inverse_A)
%   #Matrix should now be in Row Eschlon form.
%   #Back solve to get Reduced Eschelon Form
%   #iterate through each row starting with the last one
%   for i in range(1,len(A)+1):
%     r = len(A)-i
%     # print("row = " + str(r))
%     #iterate through later rows to pull from
%     for k in range(r+1,len(A)):
%       row_multiplier = copy.deepcopy(A[r][k])
%       #update each column
%       for c in range(len(A[0])):
%         # print("row multiplier = " + str(row_multiplier))
%         A[r][c] -= A[k][c]*row_multiplier
%         inverse_A[r][c] -= inverse_A[k][c]*row_multiplier
%   return inverse_A

\end{document}